\documentclass[a4paper,showpacs,superscriptaddress,twocolumn]{revtex4}
\usepackage[english]{babel}
\usepackage{graphicx}
\usepackage{xspace}
\usepackage{fancybox}
\usepackage{epic}
\usepackage{calc}
\usepackage{ifthen}
\usepackage{float}
\usepackage{color}
\usepackage{amsmath,amssymb}
\usepackage{rotating}
\usepackage{subfigure}
\usepackage{exscale}
\usepackage{setspace}
\usepackage[latin1]{inputenc}
\begin{document}

\title{Magnetism in Re-based ferrimagnetic
double perovskites}

\author{A.~Winkler}
\affiliation{Technische Universit\"{a}t Darmstadt,
Petersenstr.~23, 64287 Darmstadt, Germany}
\author{N.~Narayanan}
\affiliation{Technische Universit\"{a}t Darmstadt,
Petersenstr.~23, 64287 Darmstadt, Germany}
\author{D.~Mikhailova}
\affiliation{Technische Universit\"{a}t Darmstadt,
Petersenstr.~23, 64287 Darmstadt, Germany}
\author{K.~G.~Bramnik}
\affiliation{Technische Universit\"{a}t Darmstadt,
Petersenstr.~23, 64287 Darmstadt, Germany}
\author{H.~Ehrenberg}
\affiliation{Technische Universit\"{a}t Darmstadt,
Petersenstr.~23, 64287 Darmstadt, Germany}
\author{H.~Fuess}
\affiliation{Technische Universit\"{a}t Darmstadt,
Petersenstr.~23, 64287 Darmstadt, Germany}
\author{G.~Vaitheeswaran}
\affiliation{Royal Institute of Technology(KTH), Brinellv\"agen
23, 10044 Stockholm, Sweden} \affiliation{ACRHEM, University of
Hyderabad, Hyderabad 500 046, India}
\author{V.~Kanchana}
\affiliation{Royal Institute of Technology(KTH), Brinellv\"agen
23, 10044 Stockholm, Sweden}
\author{F.~Wilhelm}
\affiliation{European Synchrotron Radiation Facility (ESRF), 6 Rue
Jules Horowitz, BP 220, 38043 Grenoble Cedex 9, France}
\author{A.~Rogalev}
\affiliation{European Synchrotron Radiation Facility (ESRF), 6 Rue
Jules Horowitz, BP 220, 38043 Grenoble Cedex 9, France}
\author{A.~Kolchinskaya}
\affiliation{Technische Universit\"{a}t Darmstadt,
Petersenstr.~23, 64287 Darmstadt, Germany}
\author{L.~Alff}
\email{alff@oxide.tu-darmstadt.de} \affiliation{Technische
Universit\"{a}t Darmstadt, Petersenstr.~23, 64287 Darmstadt,
Germany}

\date{received 29 April 2009}

\begin{abstract}

We have investigated spin and orbital magnetic moments of the Re
$5d$ ion in the double perovskites $A_2$FeReO$_6$ ($A=$ Ba, Sr,
Ca) by X-ray magnetic circular dichroism (XMCD) at the Re
$L_{2,3}$ edges. In these ferrimagnetic compounds an unusually
large negative spin and positive orbital magnetic moment at the Re
atoms was detected. The presence of a finite spin magnetic moment
in a 'non-magnetic' double perovskite as observed in the double
perovskite Sr$_2$ScReO$_6$ proves that Re has also a small, but
finite {\em intrinsic} magnetic moment. We further show for the
examples of Ba and Ca that the usually neglected alkaline earth
ions undoubtedly also contribute to the magnetism in the
ferrimagnetic double perovskites.
\end{abstract}
\maketitle

\section{Introduction}

Ordered double perovskites of the composition $A_2 MN\textrm{O}_6$
(with $A$ an alkaline earth, $M$ a magnetic transition metal ion,
and $N$ a non-magnetic ion) have come again into the focus of
research because of their interesting magnetic properties. First,
in Sr$_2$FeMoO$_6$ a large room-temperature magnetoresistance was
observed \cite{Kobayashi:98}. Second, within the group of
ferrimagnetic double perovskites materials with higher
Curie-temperatures, $T_\textrm{C}$, than in the simple perovskites
(e.g.~doped manganites) can be obtained. At the moment the highest
$T_\textrm{C}$ values have been reported for Sr$_2$CrReO$_6$
($T_\textrm{C}\approx635$\,K)
\cite{Kato:02,Asano:04,Majewski:05apl} and Sr$_2$CrOsO$_6$
($T_\textrm{C}\approx725$\,K)
\cite{Krockenberger:07,Krockenberger:07jmmm,Lee:08}. Third, the
mechanism leading to magnetic coupling is believed to be
associated with a strong tendency to a half-metallic nature of the
charge carriers at the Fermi level
\cite{Sarma:00,Fang:01,Kanamori:01}. Clearly, these materials are
interesting candidates for spintronic applications
\cite{Serrate:07}, in particular when having in mind fully
epitaxial structures based on perovskite materials.

Recently, Majewski {\em et al.} and Sikora {\em et al.} have
proposed a simple scaling law between the Curie-temperature and
the induced magnetic moment at the non-magnetic site in the double
perovskite structure
\cite{Majewski:05apl,Majewski:05prb,Sikora:06}. Philipp {\em et
al.} have discussed that a high Curie-temperature is associated
with a tolerance factor close to one for the corresponding crystal
\cite{Philipp:03}. The only exception for this rule is found in
the series $A_2$FeReO$_6$ ($A=$ Ba, Sr, Ca). In this particular
FeRe-system it is the strongly monoclinically distorted Ca-based
compound having an anomally high $T_\textrm{C}$, namely about
540\,K \cite{Westerburg:02,Kato:02b,Michalik:07} (comparing to
about 400\,K for Sr$_2$FeReO$_6$ \cite{Kato:02b} and 325\,K for
Ba$_2$FeReO$_6$ \cite{Prellier:00,Azimonte:07}). The dimensionless
tolerance factor, $f$, in $A_2$FeReO$_6$ whose deviation from
unity implies structural distortion varies from about $f=1.057$
for $A=$ Ba over $f=0.997$ for $A=$ Sr to $f=0.943$ for $A=$ Ca
\cite{Philipp:03}. In general, the Ba-based ferrimagnetic double
perovskites are close to a structural transition into a hexagonal
lattice where ferro(i)magnetism is not allowed for symmetry
reasons; the Sr-based compounds are always close to a perfect
cubic structure with maximal $T_\textrm{C}$, and the Ca-based
double perovskites are orthorhombically or monoclinically
distorted, with still a large but - due to the reduced exchange -
clearly reduced ferrimagnetic transition temperature. The
exceptional large $T_\textrm{C}$ of Ca$_2$FeReO$_6$ is accompanied
by an insulating state at low temperatures, in contrast to
Sr$_2$FeReO$_6$ or even the similarly monoclinically distorted
Ca$_2$FeMoO$_6$ which both are metallic \cite{Iwasawa:05}. The
metal-insulator transition in Ca$_2$FeReO$_6$ has been reported to
occur between 100 and 150\,K
\cite{Westerburg:02,Kato:02b,Iwasawa:05}. This behavior has been
attributed to strongly enhanced electron-electron correlations on
the Re site due to a reduced transfer integral between Fe and Re
corresponding to an extremely large effective Coulomb repulsion,
$U_{\textrm{eff}}$, of about 4\,eV on both ions \cite{Iwasawa:05}.
This, however, is in some contradiction to the observed high
Curie-temperature, which is believed to be a consequence of a
kinetic energy gain due to the hybridization of the Fe $3d$ and Re
$5d$ $t_{\textrm{2g}}$-orbitals. The prediction that a decreased
band-filling is favorable for $T_\textrm{C}$
\cite{Chattopadhyay:01}, which could be used to conciliate a high
$T_\textrm{C}$ with a reduced Re-Re overlap, has turned out to be
not valid: an increased band-filling actually leads to a strong
$T_\textrm{C}$ enhancement for both the FeMo-system
\cite{Navarro:01} and the CrW-system \cite{Gepraegs:05}. Within
the kinetically driven exchange model
\cite{Sarma:00,Fang:01,Kanamori:01}, the increase of
$T_\textrm{C}$ is more naturally explained as a consequence of
increased band-filling. Note, that the cases of Ca$_2$FeReO$_6$
and Sr$_2$CrOsO$_6$, both being insulating and having a high
$T_\textrm{C}$ at the same time, are completely different: In the
case of Sr$_2$CrOsO$_6$ having only a tiny rhombohedral
distortion, the Os $5d$ $t_{2\text{g}}$ band is completely filled,
while for Ca$_2$FeReO$_6$ it is the structural distortion that
drives the metal-insulator transition. Recently, it was suggested
that in double perovskites with heavy ions as Re a large orbital
contribution to the magnetic moment leads to an enhanced total
magnetization above the integer value that is expected for a
half-metallic material \cite{DeTeresa:07}. This elsewhere
predicted and calculated \cite{Vaitheeswaran:05} strong influence
of spin-orbit coupling leads to a quasi-half metallicity which
still is from the view point of applications in spintronics very
high (above 90\%). Another point of interest is the possibility of
an {\em intrinsic} enhancement of the Re spin magnetic moment due
to the peculiar Re$^{5+}$ state in the ferrimagnetic double
perovskites. In this study, we present the XMCD analysis of the
system $A_2$FeReO$_6$ ($A=$ Ba, Sr, Ca), compare the experimental
data to theoretical predictions calculated within the
full-potential linear muffin-tin orbital method (FP-LMTO)
\cite{Wills:00} with included spin-orbit coupling, and complete
the so far suggested scaling law \cite{Majewski:05apl} by using
the identical method to extract separately spin and orbital
magnetic moments. Furthermore, we search for a contribution of the
alkaline earth element to the magnetic behavior, and also look for
an {\em intrinsic} Re moment in a suited double perovskite
compound with $M$ being a non-magnetic ion: Sr$_2$ScReO$_6$.

\section{Experimental}

\begin{table}
\caption{\label{tab:samp} Summary of sample properties from x-ray
diffraction at 300\,K (calculated by Rietfeld-refinement) and
SQUID magnetometry.}
\begin{center}
\begin{tabular}{l|c|c|c|c}
  material & $T_{\textrm{C}}$ & symm. & lattice & antisites \\
           &       (K) &                     & ({\AA}) & (\%)\\
  \hline\hline
Ba$_2$FeReO$_6$ & 317 & Fm$\overline{3}$m & $a=8.0571(2)$ & 0.9 \\
\hline Sr$_2$FeReO$_6$ & 418 & Fm$\overline{3}$m & $a=7.8752(4)$ &
2.6
\\ \hline
                &     &                   & $a=5.3992(2)$ &    \\
Ca$_2$FeReO$_6$ & 556 & P2$_1$/n          & $b=5.5269(2)$ & 3.6\\
                &     &                   & $c=7.6826(3)$ &    \\ \hline
                &     &                   & $a=5.6760(2)$ &    \\
Sr$_2$ScReO$_6$ &  -  &    P2$_1$/n       & $b=5.6534(2)$ & 7  \\
                &     &                   & $c=7.9862(3)$ &    \\
\end{tabular}
\end{center}
\end{table}

A summary of the sample properties is given in
Table~\ref{tab:samp}. All values where a comparison can be made to
literature values are in good agreement with these data
\cite{Westerburg:02,Kato:02}. Note that the small amount of
antisite disorder does not affect our results. The XMCD
measurements on the Re $L_{2,3}$ edges were performed at the
European Synchrotron Radiation Facility (ESRF) at beam line ID12
\cite{Rogalev:01}. The spectra were recorded within the total
fluorescence yield detection mode. The XMCD spectra were obtained
as direct difference between consecutive XANES scans (X-ray
Absorption Near Edge Spectrum) recorded with opposite helicities
of the incoming x-ray beam. To ensure that the XMCD spectra are
free from any experimental artefacts the data were collected for
both directions of the applied magnetic field of 6\,T (parallel
and antiparallel to the x-ray beam). The degree of circular
polarization of the monochromatic x-ray beam was 98\%. The
measurements were performed at about 10\,K for all samples ($T\ll
T_{\text{C}}$), if not indicated otherwise. Since the samples
measured in backscattering geometry were very thick, the spectra
were first normalized to the edge jump of unity and then corrected
from self-absorption effects.  The edge jump intensity ratio
$L_3/L_2$ was then normalized to 2.19/1 \cite{Wilhelm:01}. This is
different from the statistical 2:1 branching ratio due to the
difference in the radial matrix elements of the 2$p_{1/2}$ to
5$d$(L$_2$) and 2$p_{3/2}$ to 5$d$(L$_3$) transitions. The XMCD
measurement as a function of applied field suggests that our
samples are closer to saturation at 6\,T as is concluded by de
Teresa {\em at al.} \cite{DeTeresa:07} from high-field SQUID
measurements. This issue has to be clarified in future by
high-field XMCD measurements.

\begin{figure}[tb]
\centering{
\includegraphics[width=0.7\columnwidth,clip=]{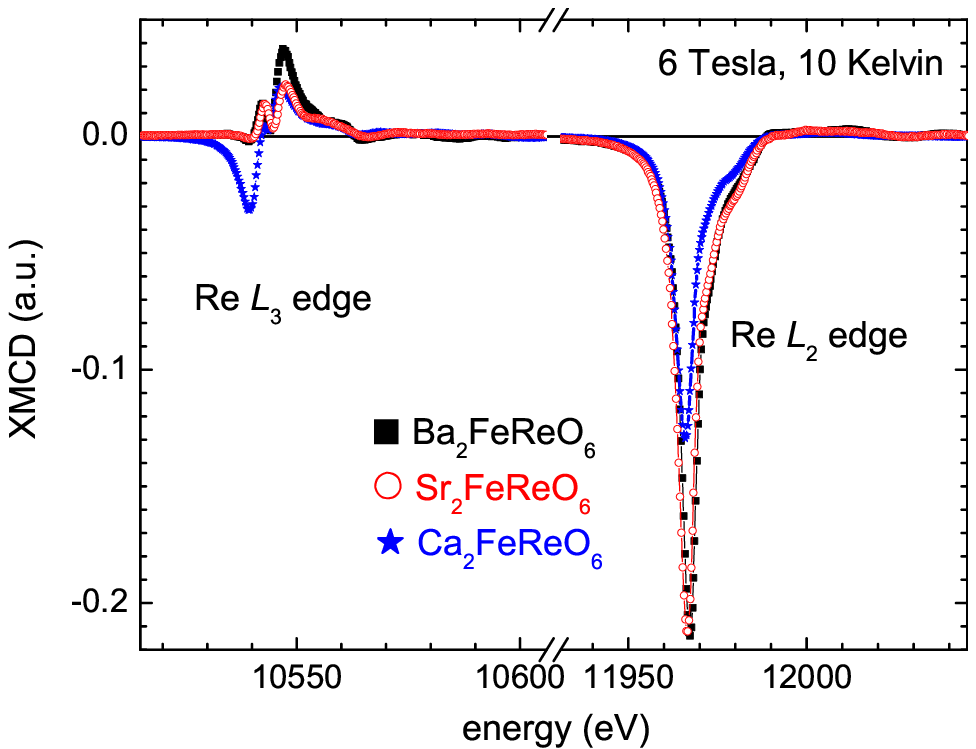}}

\caption{XMCD spectra for $A_2$FeReO$_6$($A=$Ba, Sr, Ca).}
\label{Fig:XMCD}
\end{figure}

\begin{table}
\caption{\label{tab:table1} Measured (exp., normalized to 5\,K)
and calculated (th., calculated within the generalized gradient
approximation including spin-orbit coupling (GGA+SO)) magnetic
moments at the Re site for different double perovskites at about
10\,K. For a detailed discussion of the applied band-structure
calculation see e.g.~\cite{Vaitheeswaran:05,Vaitheeswaran:06}.
Calculation in \cite{Jeng:03} is GGA {\em with} spin-orbit
coupling. The number of $d$-holes was taken from the
band-structure calculation. In our case this number was around
$5.3$. The error of the measured values is estimated as 2.5\%.}
\begin{center}
\begin{tabular}{llccc}
 & material & $m_{\textrm{S}}$ ($\mu_{\textrm{B}}$/f.u.) & $m_{\textrm{L}}$ ($\mu_{\textrm{B}}$/f.u.) & $|m_{\textrm{L}}/m_{\textrm{S}}|$
 \\ \hline

exp. & Ba$_2$FeReO$_6$ & $-0.56$ & $0.15$ & $0.27$     \\
     & Sr$_2$FeReO$_6$ & $-0.74$ & $0.21$ & $0.28$    \\
     & Ca$_2$FeReO$_6$ & $-0.47$ & $0.16$ & $0.34$\\
     & Sr$_2$CrReO$_6$ \cite{Majewski:05apl}& $-0.68$ & $0.25$ & $0.37$     \\
     & Sr$_2$ScReO$_6$ (80\,K) & $0.013$ & $-0.002$ & $0.15$     \\ \hline
th.  & Ba$_2$FeReO$_6$ & -0.65 & 0.19 & 0.29 \\
     & Sr$_2$FeReO$_6$ & -0.68 & 0.15 & 0.22\\
     & Sr$_2$FeReO$_6$ \cite{Jeng:03} & -0.85 & 0.23 & 0.27\\
     & Sr$_2$CrReO$_6$ \cite{Vaitheeswaran:05}  & -0.85 & 0.18 & 0.21\\
 \end{tabular}
\end{center}
\end{table}

\section{Results and discussion}

In this paper, the XANES spectra themselves are not further
discussed. As shown in Fig.~\ref{Fig:XMCD}, for FeRe-compounds at
both absorption edges we find a rather intense XMCD signal. This
is a clear evidence for the existence of a magnetic moment at the
Re $5d$ shell. For all three compounds, the XMCD spectra at the
$L_2$ edge are largest (as expected for $m=1$ orbitals) and
similar in shape. In Ca$_2$FeReO$_6$ the size of the XMCD signal
is by a factor of 2 smaller compared to the two other FeRe
compounds. At the $L_3$ edge, the Ca-based double perovskite again
stands out by a pronounced peak with negative XMCD signal which is
absent for Sr$_2$FeReO$_6$ and Ba$_2$FeReO$_6$. The data at the
$L_3$ edge look slightly different in amplitude as compared to
previously published data \cite{Sikora:06}. However, the data are
consistent in that the integrated XMCD intensity at the $L_3$ edge
is negative only in the case of Ca$_2$FeReO$_6$. In this sense,
all data support the unusual behavior of Ca$_2$FeReO$_6$, which
cannot only be attributed to the different ionic size of the $A$
site ions.

\begin{figure}[tb]
\centering{
\includegraphics[width=0.7\columnwidth,clip=]{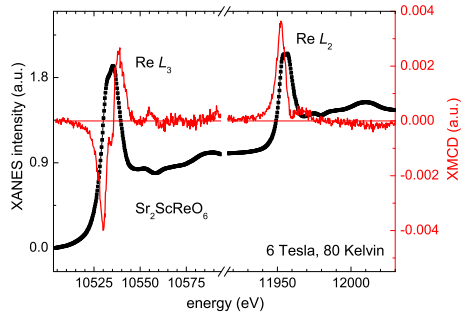}}
 \caption{XANES and derived XMCD spectra at the Re $L_2$ and $L_3$ edges
 of Sr$_2$ScReO$_6$.}
 \label{XMCD:Sc}
\end{figure}

In Fig.~\ref{XMCD:Sc} we show XANES and XMCD spectra for the
compound Sr$_2$ScReO$_6$. This compound is important because the
absence of any free electrons at Sc$^{3+}$ which has a $3d^0$
configuration will lead to a complete breakdown of the {\em
induced} magnetic moment at the Re site. This compound therefore
allows the measurement of the {\em intrinsic} magnetic moment of
Re$^{5+}$ (also in contrast to Re$^{6+}$ compounds as
Sr$_2$MgReO$_6$). Previously, Kato {\em et al.} have calculated
from a Curie-Weiss fit to the susceptibility an effective magnetic
moment of Re in Sr$_2$ScReO$_6$ of about
1.1\,$\mu_{\text{B}}/f.u.$, as expected within the ionic picture
\cite{Kato:04}. In contrast, our data show the existence of a much
smaller, but finite intrinsic moment at the Re site, indicating an
increased tendency to magnetic ordering of Re$^{5+}$. Since this
moment is present above the antiferromagnetic transition
temperature, it is not related to spin glass behavior. The spin
magnetic moment is about 50 times smaller than corresponding
induced moments on Re$^{5+}$, and the orbital magnetic moments
even by a factor of 100. However, due to the high sensitivity of
the set-up at ESRF, one can unambiguously prove the existence of
this moment. In contrast to the opposite sign of the induced
magnetic moment with respect to the applied field, the spin
magnetic moment at the Re in Sr$_2$ScReO$_6$ is aligned with the
field. This is expected because the kinetic exchange via fully
polarized spin down is not at work. This intrinsic moment of
Re$^{5+}$ therefore has to be considered as an indicator of the
tendency to unusually high magnetization of Re based double
perovskites.

\begin{figure}[tb]
\centering{%
\includegraphics[width=0.7\columnwidth,clip=]{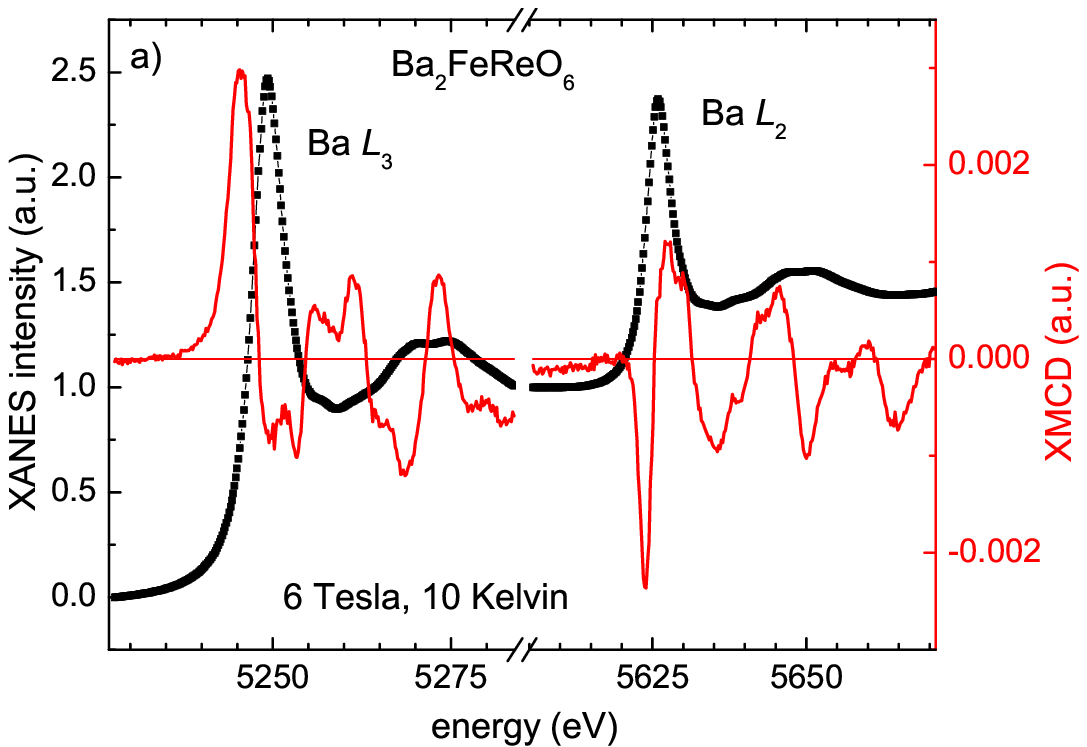}
\includegraphics[width=0.7\columnwidth,clip=]{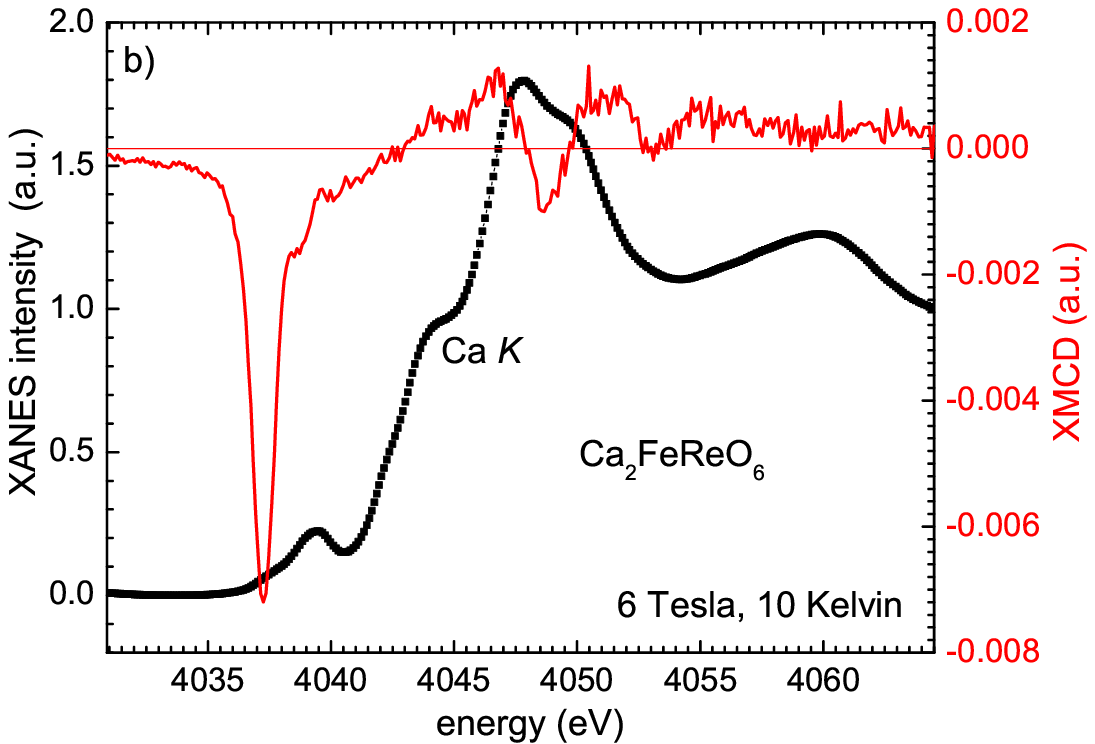}}
 \caption{XANES and derived XMCD spectra at the a) Ba $L_2$ and $L_3$ edges
 of Ba$_2$FeReO$_6$ and b) at the Ca $K$ edge of Ca$_2$FeReO$_6$.}
 \label{XMCD:Ba}
\end{figure}

As a last point, we address magnetism in the earth alkaline ions
itself, which usually are completely neglected in the magnetic
scenario. The XANES and XMCD spectra of the Ba $L_2$ and $L_3$
edges of Ba$_2$FeReO$_6$ and of the Ca $K$-edge of Ca$_2$FeReO$_6$
are shown in Fig.~\ref{XMCD:Ba}. The $5d$ spin magnetic moment
(calculated with 9 as the number of $d$-holes corresponding to the
band-structure calculation) of Ba is $\mu_{\text{S}}=-0.0065$ and
the $5d$ orbital magnetic moment $\mu_{\text{L}}=-0.0013$ (both in
$\mu_{\text{B}}/$f.u.),
$|\mu_{\text{L}}/\mu_{\text{S}}|\approx0.2$. The theoretical
predictions calculated as described elsewhere
\cite{Vaitheeswaran:05,Vaitheeswaran:06} are
$\mu_{\text{S}}=-0.0084$ and $\mu_{\text{L}}=-0.0014$ which is in
fair agreement with our experimental data. For Ca$_2$FeReO$_6$ we
can only qualitatively say that a finite magnetic moment is
observed, because the $K$-edge probes only the $4p$ orbital
magnetism. Since the $L$ edges are experimentally not accessible,
a quantative analysis cannot be done. The observation of a
magnetically polarized density of states gives clear evidence for
a magnetic interaction of the earth alkaline ions with the other
ions. The magnetic contribution of Ba in this case is a factor of
2 smaller than the contribution of the intrinsic Re moment.
Naturally, one expects that the magnetic contribution increases
with ionic size due to the increased exchange with the neighboring
ions. The clear orbital contribution in Ba$_2$FeReO$_6$ is not
unexpected due to the heavy ionic mass. Our data provide a test
for a detailed theoretical study of the magnetism in the double
perovskites, and underlines the importance of taking spin-orbit
coupling into account. Note, that for example in CrO$_2$, where
the importance of oxygen in the magnetic mechanism is undoubted,
comparable values of spin and orbital moments of the oxygen ion
have been measured \cite{Huang:2002} as compared to our results on
Ba in Ba$_2$FeReO$_6$.

In Table~\ref{tab:table1} we summarize our results for the spin
and orbital magnetic moments at the Re site as derived from the
XMCD measurements by applying the standard sum rules
\cite{Thole:92,Carra:93} and compare them to theoretical values.
Also, the ratio $\mid m_{\textrm{L}}/m_{\textrm{S}}\mid$ is
calculated, since this quantity is not affected by possible
uncertainties in the calculated number of holes. In general, the
calculated data are in surprisingly good agreement with the
measured data. One of the main reasons certainly is, that
spin-orbit coupling is taken into account from the beginning.
Note, that in the hard x-ray range the sum rules apply with high
validity due to the large spin-orbit splitting of the core level.

Let us finally discuss again our data for the three FeRe-based
compounds. Our data are in good qualitative and quantitative
agreement with literature data with one exception:
Ca$_2$FeReO$_6$. While Sikora {\em et al.} \cite{Sikora:06} find,
that the spin magnetic moment of Re in Ca$_2$FeReO$_6$ scales with
the high $T_\textrm{C}$, in our case it has the lowest spin
magnetic moment, letting Ca$_2$FeReO$_6$ stand out from the
scaling law \cite{Majewski:05apl,Majewski:05prb,Sikora:06} which
so far holds in all other cases. This behavior is certainly more
natural, since one expects that a reduced exchange will also lead
to a reduced spin magnetic moment on the Re site. Note that the
ratio of orbital and spin magnetic moments are consistent with the
previous data. As suggested previously by Kato {\em et al.}
\cite{Kato:02b}, a Re $t_{\text{2g}}$ orbital ordered state or its
glass-state analog associated with the monoclinic lattice
distortion occurs, pointing out the importance of correlation
effects in this compound. Recently, Sikora {\em et al.}
\cite{Sikora:09} proposed a scenario with a complex competition
between two phases with different electronic and crystallographic
structure. Our data give further indication that Ca$_2$FeReO$_6$
is exceptional among the double perovskites due to the strong
octahedral-site distortions.

\section{Summary}

In summary, we have elucidated the Re magnetic moments in the
FeRe-based series of double perovskites as a function of the earth
alkaline ion, confirming the exceptional position of
Ca$_2$FeReO$_6$. We have measured a finite {\em intrinsic}
magnetic moment at the Re$^{5+}$ site in Sr$_2$ScReO$_6$
indicating the tendency to enhanced magnetic moments observed in
Re based double perovskites. Furthermore, for the first time we
were able to measure by XMCD the magnetic moments directly at the
alkaline earth site itself. Our result shows that the usually
neglected Ca and Ba ions play a role in the magnetic scenario of
the kinetically driven exchange model, comparable in size to the
role of oxygen.

This work was supported by the ESRF (HE-2114/2115/2379).

\end{document}